\documentclass[aps,twocolumn,prl,superscriptaddress,amsmath,amssymb]{revtex4-1}

\usepackage{times}
\usepackage{amsmath}
\usepackage{amsfonts}
\usepackage{amssymb}
\usepackage{dcolumn}
\usepackage{bm}
\usepackage{txfonts}
\usepackage{ulem}
\usepackage{braket}
\usepackage[dvipdfmx]{graphicx}
\usepackage[usenames]{xcolor}

\def\journal #1#2#3#4{{#1} {\bf #2}, #3 (#4).}

\begin{document}
\title{Antiferromagnetic Kitaev Interactions in Polar Spin-Orbit Mott Insulators}

\author{Yusuke Sugita}
\affiliation{Department of Applied Physics, The University of Tokyo, Bunkyo, Tokyo 113-8656, Japan}
\author{Yasuyuki Kato}
\affiliation{Department of Applied Physics, The University of Tokyo, Bunkyo, Tokyo 113-8656, Japan}
\author{Yukitoshi Motome}
\affiliation{Department of Applied Physics, The University of Tokyo, Bunkyo, Tokyo 113-8656, Japan}

\date{\today}
\begin{abstract}
A bond-directional anisotropic exchange interaction, called the Kitaev interaction, is a promising route to realize quantum spin liquids.
The Kitaev interactions were found in Mott insulators with the strong spin-orbit coupling, in the presence of quantum interference between indirect electron transfers.
Here we theoretically propose a different scenario by introducing a polar structural asymmetry that unbalances the quantum interference.
We show that the imbalance activates additional exchange processes and gives rise to a dominant antiferromagnetic Kitaev interaction, in stark contrast to the conventional ferromagnetic ones. 
We demonstrate by {\it ab initio} calculations that polar Ru trihalides with multiple anions, $\alpha$-RuH$_{3/2}X_{3/2}$ ($X$=Cl and Br), exhibit the antiferromagnetic Kitaev interaction whose magnitude is several times larger compared to existing candidates.
Our proposal opens the way for materializing the Kitaev spin liquids in unexplored parameter regions.
\end{abstract}
\maketitle

Entanglement between spin and orbital degrees of freedom of electrons is a fertile source of exotic phases of matter.
Such entanglement typically appears via the strong spin-orbit coupling (SOC) in the valence shells of heavy atomic elements as well as strong electron interactions.
In particular, in transition metal compounds with 4$d$ and 5$d$ electrons, synergy between the SOC and electron interactions brings about intriguing quantum phases, such as Weyl semimetals and topological Mott insulators~\cite{PhysRevB.83.205101,doi:10.1146/annurev-conmatphys-020911-125138}.
Among them, the strongly correlated regime, the so-called spin-orbit Mott insulator, has been extensively studied as a key for realizing exotic magnetism, such as noncollinear magnetic ordering and quantum spin liquids (QSLs)~\cite{doi:10.1146/annurev-conmatphys-020911-125138,doi:10.1146/annurev-conmatphys-031115-011319}.

In the spin-orbit Mott insulators, the SOC and electron interactions give rise to anisotropic exchange interactions between the localized electrons. 
Among various types of the anisotropic interactions, bond-directional Ising-type interactions have attracted great interest for over a decade, as a clue for realizing the celebrated Kitaev model that provides an exact QSL in the ground state~\cite{Kitaev2006}.
It was pointed out that the Kitaev interactions are realized under two requisites~\cite{PhysRevLett.102.017205}: (i)~a Kramers doublet with the effective total angular momentum $j_{\rm eff}=1/2$ under the cubic crystalline electric field (CEF) and the SOC and (ii)~quantum interference between the indirect electron transfers via different $90^\circ$ cation-ligand-cation bonds.
These are approximately satisfied in some 4$d$ and 5$d$ transition metal compounds with the low-spin $d^5$-electron configuration and the edge-sharing honeycomb structure of ligand octahedra, such as $A_2$IrO$_3$ ($A$=Na, Li)~\cite{PhysRevB.82.064412,PhysRevLett.108.127203} and $\alpha$-RuCl$_3$~\cite{PhysRevB.90.041112,PhysRevB.91.094422}.
Indeed, recent theoretical and experimental studies revealed dominant ferromagnetic (FM) Kitaev interactions in these magnets~\cite{PhysRevLett.105.027204,PhysRevLett.110.097204,PhysRevLett.112.077204,1367-2630-16-1-013056,PhysRevLett.113.107201,PhysRevB.93.214431,Winter_2017,HwanChun2015,Banerjee2016,PhysRevLett.118.187203,PhysRevB.98.100403}.
Despite the parasitic magnetic orders at low temperature presumably due to other subdominant interactions, anomalous behaviors, potentially ascribed to the proximity to the Kitaev QSL, have been reported above the transition temperature and the critical magnetic field~\cite{PhysRevLett.114.147201,Nasu2016,Banerjee2016,PhysRevB.96.041405,Do2017,PhysRevLett.118.187203,PhysRevLett.118.107203,PhysRevLett.119.037201,PhysRevB.98.100403,PhysRevLett.120.217205,Kasahara2018}.

While the Kitaev QSLs have received keen attention in the field of not only magnetism but also quantum computation~\cite{Kitaev2006,KITAEV20032}, the candidate materials are still limited.
Recently, several efforts have been made to extend the candidates. 
For instance, the $d^7$ electron configuration in the high-spin state~\cite{PhysRevB.97.014407,PhysRevB.97.014408} and the $f$-electron multiplets~\cite{PhysRevB.95.085132,PhysRevB.98.054408,Jang2018,Luo2019,Xing2019} were nominated for alternative $j_{\rm eff}=1/2$ Kramers doublets in the requisite~(i).  
In addition, the networks with parallel-edge-sharing octahedra~\cite{PhysRevB.86.140405,PhysRevB.91.155135} and organic ligand bridges~\cite{PhysRevLett.119.057202} were proposed as alternative ligand geometries in the requisite~(ii). 
These point out interesting possibilities of the Kitaev candidates, but such challenges have been just initiated and await for further experimental verifications.
Moreover, recent theoretical studies predict intriguing QSL phases in a magnetic field when the Kitaev interactions are antiferromagnetic (AFM)~\cite{PhysRevB.83.245104,PhysRevB.97.241110,PhysRevB.98.014418,PhysRevB.98.060416,hickey2019emergence,Ronquillo2018}, but there are a few proposals for the realization~\cite{PhysRevB.98.054408,Jang2018}. 

In this work, we theoretically propose an alternative scenario to realize the Kitaev interactions.
We find that a polar crystalline structure, which unbalances the quantum interference in the requisite~(ii), gives rise to an AFM Kitaev interaction.
We show that it originates from different perturbation processes from the conventional mechanism, which are activated by the polar imbalance. 
In order to estimate the exchange coupling constants quantitatively, we perform {\it ab initio} calculations for candidate polar materials with multiple anions.
We find that Ru trihalides with hydrides H$^-$ potentially exhibit dominant AFM Kitaev interactions, whose magnitude is considerably larger than the conventional ferromagnetic one.

\begin{figure}[t]
\centering
\includegraphics[width=1\columnwidth]{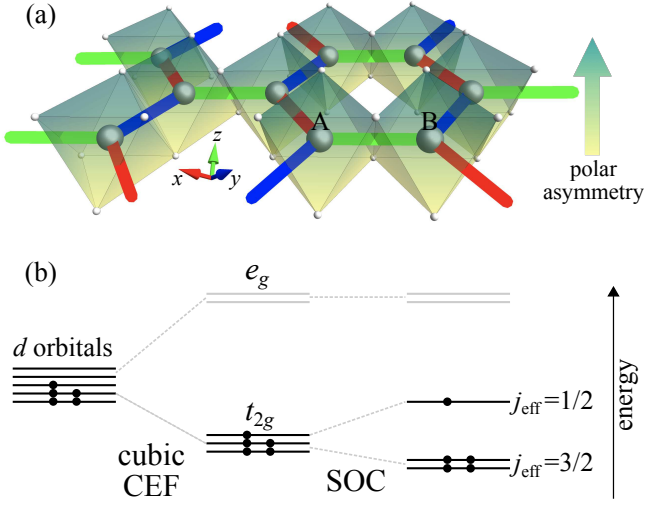}
\caption{
(a)~Schematic picture of a monolayer of polar honeycomb-layered transition metal compounds.
The large and small spheres represent the transition metal cations and the ligand ions, respectively.
The color gradation of the octahedra depicts polar asymmetry in the perpendicular direction to the honeycomb plane.
The red, blue, and green bonds denote the $x$, $y$, and $z$ bonds, respectively [see Eq.~\eqref{hop}].
The orthogonal $xyz$ axes are taken along the directions from a cation to the surrounding ligands in the ideal octahedron.  
The labels A and B  indicate two sublattices of the honeycomb structure.
(b)~Schematic energy levels of the low-spin $d^5$-electron configuration under the cubic CEF and the SOC.
The black dots indicate $d$ electrons.
}
\label{setup}
\end{figure}

We begin with a multiorbital Hubbard model for the low-spin $d^5$ state on a polar honeycomb-layered structure.
The honeycomb layer is composed of an edge-sharing network of the ligand octahedra as demonstrated later [Fig.~\ref{setup}(a)].
We assume polar asymmetry in the perpendicular direction to the honeycomb plane, which would be realized, e.g., at surfaces and interfaces, and by arranging different ligands in the upper and lower triangles of the octahedra.    
We assume threefold rotational symmetry around the [111] axis through every cation site and mirror symmetry with respect to the plane spanned by the [001] and [110] axes through the center of nearest-neighbor bonds.
In this situation, the $d$ levels are split into the $e_g$ and $t_{2g}$ manifolds by the dominant cubic CEF, and the five electrons occupy the $t_{2g}$ levels in the low-spin state, as shown in the middle panel in Fig.~\ref{setup}(b). 
We consider the multiorbital Hubbard model for holes in the $t_{2g}$ orbitals, whose Hamiltonian consists of four terms as $H = H_{\rm hop} + H_{\rm int} + H_{\rm SOC} + H_{\rm tri}$~\cite{PhysRevB.93.214431}.

The first term $H_{\rm hop}$ describes the kinetic energy of holes.
We here take into account the transfer integrals between nearest-neighbor cations only~\cite{SM}. 
$H_{\rm hop}$ is written in the matrix form of 
\begin{equation}
H_{\rm hop}
=
- 
\sum_{\langle ij \rangle}
\bm{c}^\dagger_i
\left(
\hat{T}_{\gamma_{ij}}
\otimes
\hat{\sigma}_0
\right)
\bm{c}_j
+{\rm h.c.}
,
\label{hop}
\end{equation}
where $\hat{T}_{\gamma_{ij}}$ denotes the transfer integrals including both direct and indirect contributions [see Eq.~\eqref{hopz} below], $\hat{\sigma}_0$ is the identity matrix, and $\bm{c}^\dagger_i = \left(c^\dagger_{i\,yz\,\uparrow}\,\,c^\dagger_{i\,yz\,\downarrow}\,\,c^\dagger_{i\,zx\,\uparrow}\,\,c^\dagger_{i\,zx\,\downarrow}\,\,c^\dagger_{i\,xy\,\uparrow}\,\,c^\dagger_{i\,xy\,\uparrow} \right)$; $c^{\dagger}_{i m \sigma}$ ($c_{i m \sigma}$) is the creation (annihilation) operator of a hole at site $i$ with orbital $d_m$ ($m$ = $yz$, $zx$, or $xy$) and spin $\sigma =\uparrow$ or $\downarrow$ (the spin quantization axis is taken along the [001] axis). 
Here, sites $i$ and $j$ belong to the A and B sublattices of the honeycomb lattice, respectively, $\left< ij \right>$ denotes nearest-neighbor pairs, and $\gamma_{ij} =x$, $y$, $z$ denotes the $\gamma_{ij}$ bond between the sites $i$ and $j$ [see Fig.~\ref{setup}(a)].
From the crystalline symmetry, the transfer integrals, for instance, on the $z$ bonds, are given by 
\begin{equation}
\hat{T}_{z}
=
\left(
\begin{array}{ccc}
t_{1}			&t_{2}-\eta_1 /2		&t_{4}+\eta_2 /2 \\
t_{2}+\eta_1 /2	&t_{1} 			&t_{4}-\eta_2 /2 \\
t_{4}-\eta_2 /2	&t_{4}+\eta_2 /2	&t_{3}
\end{array}
\right).
\label{hopz}
\end{equation}
When the system is nonpolar, $\eta_1$ and $\eta_2$ both vanish. 
Furthermore, $t_4$ and $\eta_2$ are small when the octahedra are not largely distorted~\cite{PhysRevLett.112.077204}. 
In such cases, the exchange processes via $t_2$ predominantly contribute to FM Kitaev interactions~\cite{PhysRevLett.102.017205,1367-2630-16-1-013056,PhysRevLett.113.107201,PhysRevB.93.214431}. 

The second term $H_{\rm int}$ denotes the onsite Coulomb interactions, which is given by 
\begin{equation}
H_{\rm int}
=
\frac{1}{2} \sum_{mnm'n'} U_{mnm'n'}
\sum_{i}
\sum_{\sigma \sigma'}
c^{\dagger}_{i m\sigma} c^{\dagger}_{i n\sigma'} c_{i n'\sigma'} c_{im'\sigma}.
\label{onsite}
\end{equation}
Assuming the rotational symmetry of the Coulomb interaction, we set $U_{mmmm} = U$,  $U_{mnmn} = U-2J_{\rm H}$, and $U_{mnnm}=U_{mmnn}=J_{\rm H}$ ($m\neq n$), where $U$ is the intraorbital Coulomb interaction and $J_{\rm H}$ is the Hund's coupling, respectively~\cite{sugano1970}.
The third and last terms in $H$ describe the SOC and the trigonal CEF splitting as 
\begin{eqnarray}
H_{\rm SOC}
&=&
- 
\frac{\lambda}{2}
\sum_{i}
\bm{c}^\dagger_i
\left(
\begin{array}{ccc}
0					&{\rm i} \hat{\sigma}_z	&-{\rm i} \hat{\sigma}_y \\
-{\rm i} \hat{\sigma}_z		&0 					&{\rm i} \hat{\sigma}_x \\
{\rm i} \hat{\sigma}_y	&-{\rm i} \hat{\sigma}_x	&0
\end{array}
\right)
\bm{c}_i, 
\label{SOC} \\
H_{\rm tri}
&=&
- 
\sum_{i}
\bm{c}^\dagger_i
\left[
\left(
\begin{array}{ccc}
0		&\Delta_{\rm tri}	&\Delta_{\rm tri} \\
\Delta_{\rm tri} 	&0 		&\Delta_{\rm tri} \\
\Delta_{\rm tri} 	&\Delta_{\rm tri}	&0
\end{array}
\right)
\otimes
\hat{\sigma}_0
\right]
\bm{c}_i,
\label{tri}
\end{eqnarray}
respectively, where $\hat{\sigma}_\alpha$ ($\alpha$=$x$, $y$, and $z$) is the Pauli matrix.

The $t_{2g}$ manifold is split into $j_{\rm eff}=1/2$ doublet and $j_{\rm eff}=3/2$ quartet under the SOC, and the ground state is given by a single-hole state in the $j_{\rm eff}=1/2$ manifold per site [Fig.~\ref{setup}(b)].
When the Coulomb interactions localize the holes to form the spin-orbit Mott insulating state~\cite{PhysRevLett.101.076402}, the low-energy physics is governed by the exchange interactions between two pseudospins describing the Kramers pair of the $j_{\rm eff}=1/2$ states. 
The effective interactions on neighboring sites can be derived by using the second-order perturbations in terms of the hopping transfers in Eq.~\eqref{hop}, which are summarized into the generic form of $H_{\rm spin}= \sum_{\langle ij \rangle} \bm{S}^{\rm T}_i \hat{J}_{\gamma_{ij}} \bm{S}_j$, where $\bm{S}_i$ denotes the pseudospin operator at site $i$.
From the crystalline symmetry, the exchange interactions $\hat{J}_{\gamma_{ij}}$, e.g., for the $z$ bonds, are written as 
\begin{equation}
\hat{J}_{z}
=
\left(
\begin{array}{ccc}
J			&D+\Gamma	&-D'+\Gamma'\\
-D+\Gamma	&J 			&D'+\Gamma'\\
D'+\Gamma'	&-D'+\Gamma'	&J+K
\end{array}
\right),
\label{ex}
\end{equation}
where $J$ is the coupling constant for the isotropic Heisenberg exchange interaction, $K$ is for the Kitaev interaction, $\Gamma$ and $\Gamma'$ are for the off-diagonal symmetric exchanges interactions, and $D$ and $D'$ are for the Dzyaloshinkii-Moriya interactions~\cite{PhysRevLett.112.077204,PhysRevB.93.214431}.  
The coupling constants for the $x$ and $y$ bonds are obtained by the threefold rotations on Eq.~\eqref{ex}.

\begin{figure}[t]
\centering
\includegraphics[width=1.0\columnwidth]{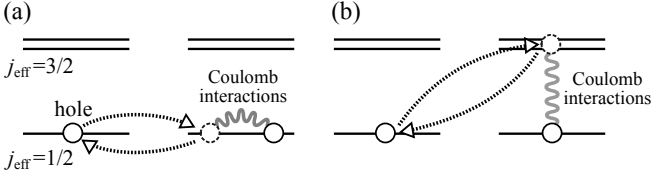}
\caption{
Perturbation processes (a) within the $j_{\rm eff}=1/2$ states and (b) via the $j_{\rm eff}=3/2$ states.
}
\label{second}
\end{figure}

When neglecting the trigonal CEF, there are only two types of perturbation processes within the $t_{2g}$ manifold contributing to the coupling constants in Eq.~\eqref{ex}: One is within the $j_{\rm eff}=1/2$ manifold [Fig.~\ref{second}(a)] and the other is via the $j_{\rm eff}=3/2$ manifold [Fig.~\ref{second}(b)]~\cite{PhysRevB.93.214431}.
We find that the polar asymmetry in the lattice structure gives a crucial contribution to the former process.
This is explicitly shown by considering the effective hopping transfers within the $j_{\rm eff}=1/2$ states. 
By projecting Eq.~\eqref{hop} onto the $j_{\rm eff}=1/2$ states, we obtain
\begin{equation}
H_{\rm hop}^{\rm eff}
=
- 
\sum_{\langle ij \rangle}
\tilde{\bm{c}}^\dagger_i
\left(
\tilde{t} 
\hat{\sigma}_0
- {\rm i} 
\tilde{\eta}
\hat{\sigma}_{\gamma_{ij}}
\right)
\tilde{\bm{c}}_j
+{\rm h.c.}
,
\label{hop1/2}
\end{equation}
where $\tilde{t} = (2t_1+t_3)/3$ and $\tilde{\eta} = \eta_1/3$; $\tilde{\bm{c}}^\dagger_i =(\tilde{c}^\dagger_{i \tilde{\uparrow}}\,\, \tilde{c}^\dagger_{i \tilde{\downarrow}})$, $\tilde{c}^\dagger_{i \tilde{\sigma}}$ ($\tilde{c}_{i \tilde{\sigma}}$) is the creation (annihilation) operator of a hole in the $j_{\rm eff}=1/2$ manifold at site $i$ with pseudospin $\tilde{\sigma} =\tilde{\uparrow}$ or $\tilde{\downarrow}$. 
For simplicity, we neglect contributions proportional to $\eta_2$ in Eq.~\eqref{hop1/2}, which are small in realistic situations (see Table~\ref{mlwf}). 

Equation~\eqref{hop1/2} shows that the effective hopping transfers include spin- and bond-dependent contributions proportional to $\tilde{\eta}$ in the presence of polar asymmetry.
This is similar to the Rashba-type SOC. 
By considering the second-order perturbation with respect to the hopping transfers in Eq.~\eqref{hop1/2} [Fig.~\ref{second}(a)], we obtain
\begin{equation}
J \sim \frac4U \left(\tilde{t}^2 - \tilde{\eta}^2\right), \ \
K \sim \frac8U \tilde{\eta}^2, \ \ 
D \sim -\frac8U \tilde{t} \tilde{\eta},
\label{JKD}
\end{equation}
and $\Gamma = \Gamma' = D' = 0$.
The full expressions including $\eta_2$ and the perturbation process via the $j_{\rm eff}=3/2$ states [Fig.~\ref{second}(b)] are given in Supplemental Material~\cite{SM}.
An important finding in Eq.~\eqref{JKD} is that the Kitaev interaction is AFM, $K>0$, and its magnitude is proportional to $1/U$. 
This is in stark contrast to the conventional scenario for nonpolar spin-orbit Mott insulators~\cite{PhysRevLett.102.017205}, where a dominant FM Kitaev interaction proportional to $J_{\rm H}/U^2$ is predicted from the perturbation process via the $j_{\rm eff}=3/2$ states [Fig.~\ref{second}(b)].

\begin{figure}[t]
\centering
\includegraphics[width=1.0\columnwidth]{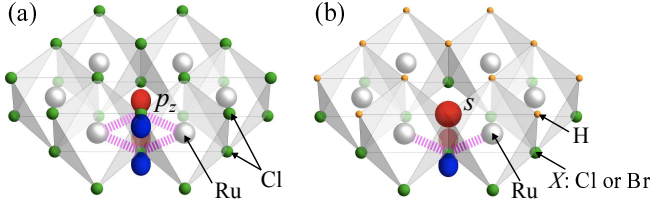}
\caption{
Schematic pictures of monolayers of (a)~nonpolar $\alpha$-RuCl$_3$ and (b)~polar $\alpha$-RuH$_{3/2} X_{3/2}$ ($X$= Cl and Br).
The dotted lines indicate indirect transfers through the ligand $p_z$ orbitals. 
The $s$ orbital in the hydride H$^-$ significantly suppresses the indirect transfers.
}
\label{polar}
\end{figure}

In reality, however, the AFM and FM Kitaev interactions can compete with each other.
In order to estimate their realistic values, we perform {\it ab initio} calculations by \textsc{openmx} code~\cite{openmx} (see Supplemental Material for the computational details~\cite{SM}).
Starting from a Kitaev candidate $\alpha$-RuCl$_3$, we consider polar asymmetric materials by introducing different anions on two ligand layers sandwiching the Ru honeycomb layer (see Fig.~\ref{polar}). 
Note that the syntheses of similar polar structures with multiple anions were reported for the layered transition metal compounds~\cite{Lu2017}. 
In particular, we focus on a monolayer form of half hydride compounds, $\alpha$-RuH$_{3/2} X_{3/2}$ ($X$= Cl and Br). 
It is worth noting that the hydrides bring about extreme asymmetry to the quantum interference in the requisite (ii): $s$ orbitals of  H$^-$ strongly suppress the indirect transfers between $t_{2g}$ orbitals from symmetry [see Fig.~\ref{polar}(b)]~\cite{kageyama2018expanding}. 

\begin{figure*}[t]
\centering
\includegraphics[width=2.0\columnwidth]{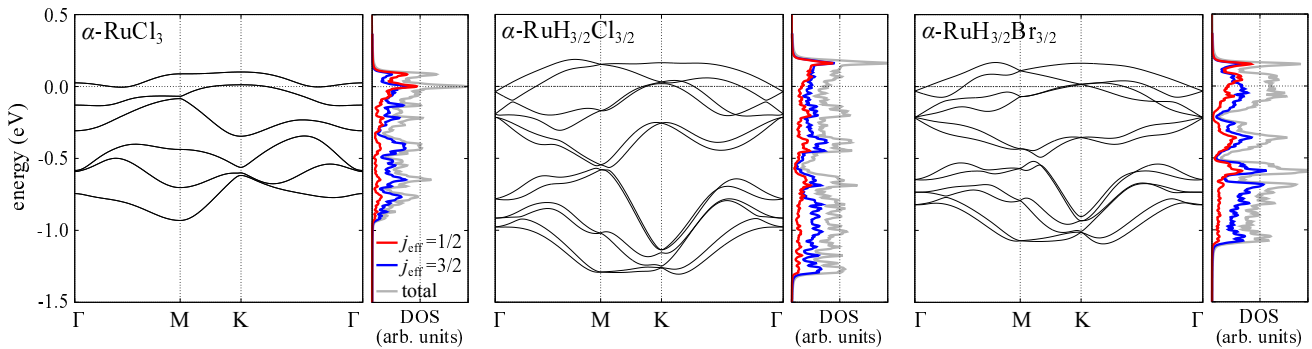}
\caption{
Electronic band structures and the density of states (DOS) for monolayers of nonpolar $\alpha$-RuCl$_3$ and polar $\alpha$-RuH$_{3/2} X_{3/2}$ ($X$= Cl and Br) obtained by the relativistic {\it ab initio} calculations.
For each case, the right panel displays the DOS: The red and blue lines indicate the projected DOS onto the $j_{\rm eff}=1/2$ and $3/2$ states, respectively, while the gray line denotes the total DOS.
The Fermi level is set to be zero in all cases. 
}
\label{band}
\end{figure*}

Figure~\ref{band} shows the electronic band structures for monolayers of nonpolar $\alpha$-RuCl$_3$ and polar $\alpha$-RuH$_{3/2}$$X_{3/2}$ ($X$= Cl and Br), obtained by the relativistic {\it ab initio} calculations for the paramagnetic state.
In all cases, the Fermi level locates in the well-isolated $t_{2g}$ manifold, and the bandwidth changes in accordance with the optimized lattice constants: 5.97~\AA, 5.35~\AA, and 5.67~\AA~for $\alpha$-RuCl$_3$, $\alpha$-RuH$_{3/2}$Cl$_{3/2}$, and $\alpha$-RuH$_{3/2}$Br$_{3/2}$, respectively.
We note that while each band of the nonpolar $\alpha$-RuCl$_3$ is twofold degenerate, the degeneracy is lifted for polar $\alpha$-RuH$_{3/2}$$X_{3/2}$.
We also show the projected density of states (DOS) onto the $j_{\rm eff}=1/2$ and $3/2$ states as well as the total DOS in Fig.~\ref{band}.
The results indicate that $\alpha$-RuH$_{3/2} X_{3/2}$ ($X$= Cl and Br) share the common trend to $\alpha$-RuCl$_3$: The $j_{\rm eff}=1/2$ state has larger contributions to the electronic states near the Fermi level than the lower-energy regions.

\begin{table}
\centering
\begin{ruledtabular}
\begin{tabular}{c|ccccccc}
							&$t_1$	&$t_2$	&$t_3$	&$t_4$	&$\eta_1$	&$\eta_2$	&$\Delta_{\rm tri}$\\
\hline
$\alpha$-RuCl$_{3}$				&$45$		&$159$		&$-117$	&$-22$		&$0$		&$0$		&$-20$\\
$\alpha$-RuH$_{3/2}$Cl$_{3/2}$	&$143$		&$-25$		&$-227$	&$-81$		&$303$		&$30$		&$-12$\\
$\alpha$-RuH$_{3/2}$Br$_{3/2}$	&$97$		&$0$		&$-128$	&$-67$		&$283$		&$23$		&$8$
\end{tabular}
\end{ruledtabular}
\caption{
Nearest-neighbor transfer integrals and trigonal CEF for monolayers of nonpolar $\alpha$-RuCl$_3$ and polar $\alpha$-RuH$_{3/2} X_{3/2}$ ($X$= Cl and Br) obtained by {\it ab initio} calculations. 
See the definitions in Eqs.~\eqref{hopz} and \eqref{tri}.
The unit is in meV. 
Further-neighbor transfers are shown in Supplemental Material~\cite{SM}. 
}
\label{mlwf}
\end{table}

By the maximally localized Wannier analysis~\cite{PhysRevB.56.12847,PhysRevB.65.035109} for the $t_{2g}$ bands, we estimate the transfer integrals in Eq.~\eqref{hopz} and the trigonal CEF splitting in Eq.~\eqref{tri}~\cite{SM}.
The estimates are summarized in Table~\ref{mlwf}.
The result for nonpolar $\alpha$-RuCl$_3$ is consistent with the previous study~\cite{PhysRevB.93.214431}. 
In the polar cases, $\eta_1$ and $\eta_2$ become nonzero as expected.
Remarkably, the most dominant $t_2$ in the nonpolar case is substantially suppressed, and $\eta_1$ becomes most dominant in both hydride compounds. 
We note that $\eta_2$ is much smaller than $t_1$, $t_3$, and $\eta_1$, and moreover, the trigonal CEF $\Delta_{\rm tri}$ remains much smaller than the empirical value of the SOC for Ru$^{3+}$, $\lambda \sim 150$~meV~\cite{PhysRevB.93.214431}; these rationalize our tight-binding analysis above.

By using the estimates in Table~\ref{mlwf}, we evaluate the exchange coupling constants for $\alpha$-RuCl$_3$ and $\alpha$-RuH$_{3/2} X_{3/2}$ ($X$= Cl and Br). 
In the calculations, we include all the perturbation processes within the $t_{2g}$ manifold and the effect of trigonal CEF splitting~\cite{SM}, and take $U=3.0$~eV, $J_{\rm H}=0.6$~eV, and $\lambda=0.15$~eV for Ru$^{3+}$~\cite{PhysRevB.93.214431}.
Table~\ref{coupling} summarizes the results.
Again, our results for $\alpha$-RuCl$_3$ well agree with the previous theoretical study~\cite{PhysRevB.93.214431}, which reported the dominant FM Kitaev interaction $K$ and the subdominant off-diagonal symmetric interaction $\Gamma$. 
We find, however, that the introduction of hydrides makes $K$ being AFM with much larger amplitudes, while suppressing $\Gamma$.
At the same time, the FM $J$ as well as $\Gamma'$ is enhanced. 
Our results indicate the exchange interactions of the polar hydride compounds are governed by the dominant AFM $K$ and the subdominant FM $J$. 
This is understood from Eq.~\eqref{JKD} with the transfer integrals in Table~\ref{mlwf}.
By the hydiride substitutions, $\eta_1$ becomes most dominant among the transfer integrals, while $t_1$ and $t_3$ are secondary and have the opposite signs, as shown in Table~\ref{mlwf}.    
Then, the exchange interactions are dominated by the terms proportional to $\tilde{\eta}^2$ in Eq.~\eqref{JKD}.
This leads to $K \sim -2J$, which approximately holds in Table~\ref{coupling}.
We note that the $t_{2g}$-$e_g$ mixing also leads to $K \sim -2J > 0$~\cite{khaliullin2005orbital,PhysRevLett.110.097204}, but the values are much smaller than our results because of the large cubic CEF splitting between the $t_{2g}$ and $e_g$ manifolds~\cite{PhysRevB.88.035107,PhysRevLett.113.107201,PhysRevB.93.214431}.

\begin{table}
\centering
\begin{ruledtabular}
\begin{tabular}{c|cccccc}
											&$J$		&$K$	&$\Gamma$	&$\Gamma'$	&$D$	&$D'$\\
\hline
$\alpha$-RuCl$_{3}$			
&$-0.8$	&$-8.0$	&$5.8$			&$-2.3$		&$0.0$		&$0.0$	\\
$\alpha$-RuH$_{3/2}$Cl$_{3/2}$	
&$-35.6$	&$69.8$	&$-1.0$		&$11.2$		&$-3.6$	&$0.0$\\
$\alpha$-RuH$_{3/2}$Br$_{3/2}$	
&$-25.9$	&$50.5$	&$2.8$			&$4.4$			&$-0.6$	&$2.5$	
\end{tabular}
\end{ruledtabular}
\caption{
Exchange coupling constants for monolayers of nonpolar $\alpha$-RuCl$_3$ and polar $\alpha$-RuH$_{3/2} X_{3/2}$ ($X$= Cl and Br) estimated by the perturbation theory by using the parameters in Table~\ref{mlwf} and $U=3.0$~eV, $J_{\rm H}=0.6$~eV, and $\lambda=0.15$~eV~\cite{PhysRevB.93.214431}.
The unit is in meV.
See Supplemental Material for further neighbors~\cite{SM}.
}
\label{coupling}
\end{table}

Finally, let us discuss the relevance of our results to the study of Kitaev QSLs.
The AFM Kitaev interactions have recently attracted great interest as they not only preserve a topological QSL phase in a broader range of magnetic fields than the FM ones but also lead to an enigmatic intermediate phase before entering the forced-ferromagnetic phase, which is not seen in the FM case~\cite{PhysRevB.83.245104,PhysRevB.97.241110,PhysRevB.98.014418,PhysRevB.98.060416,hickey2019emergence,Ronquillo2018}.
Our results indicate that the AFM Kitaev interactions can be obtained simply by introducing the polar asymmetry in existing candidates. 
While our hydride compounds bring the subdominant FM Heisenberg interactions, which may stabilize the zigzag AFM order in the ground state~\cite{PhysRevLett.110.097204,PhysRevLett.112.077204}, further tuning of the electronic parameters, not only by substituting different anions but also, e.g., by making surfaces and interfaces, would be helpful to study the intriguing physics of Kitaev QSLs in unexplored parameter regions.

To summarize, we have theoretically uncovered an alternative mechanism to realize the Kitaev interactions in the spin-orbit Mott insulator by introducing polar asymmetry in the honeycomb-layered structure.
We showed that the perturbation processes through asymmetric indirect transfers between the $j_{\rm eff}$=1/2 states give rise to the AFM Kitaev interactions, which compete with the conventional FM ones originating from the perturbation processes via the $j_{\rm eff}=3/2$ states. 
We confirmed our scenario by {\it ab initio} calculations and proposed that a family of polar Ru halides, $\alpha$-RuH$_{3/2}X_{3/2}$ ($X$= Cl and Br), are good candidates for realizing the dominant AFM Kitaev interactions of several tens meV.
The scenario based on polar asymmetry is generic, not limited to the materials with multiple anions but extended to surfaces and interfaces of layered transition metal compounds. 
It also points to a possibility of tuning the Kitaev interaction by applying an electric voltage.
Our results would provide a unique step towards crystallographic, structural, and electronic designing of the Kitaev magnets.

\begin{acknowledgments}
Y.S. thanks Youhei Yamaji for constructive suggestions and helpful comments.  
Y.S. was supported by the Japan Society for the Promotion of Science through a research fellowship for young scientists and the Program for Leading Graduate Schools (MERIT).
This research was supported by Grant-in-Aid for Scientific Research under Grants No.~JP16H02206 and JP18K03447 and JST CREST (JPMJCR18T2). 
\end{acknowledgments}


\clearpage
\setcounter{figure}{0}
\setcounter{equation}{0}
\setcounter{table}{0}
\renewcommand{\thefigure}{S\arabic{figure}}
\renewcommand{\theequation}{S\arabic{equation}}
\renewcommand{\thetable}{S\arabic{table}}                                                      

\begin{center}                                                    
{\bf ---Supplemental Material---}                               
\end{center}
\subsection*{S1.~~Exchange coupling constants in the absence of trigonal crystalline electric field}
When the trigonal CEF splitting is absent, one can obtain the exchange coupling constants in Eq.~(6) analytically for all the perturbation processes in the $t_{2g}$ manifold.  
The expressions are given by~\cite{SM.PhysRevB.93.214431}   
\begin{align}
J
=&
\frac{4 \mathbb{A}}{9}
\left\{
(2 t_1 + t_3 )^2 - \eta^2_1
\right\}
\nonumber \\
&-\frac{2\mathbb{B}}{9}
\left\{
8 (t_1 - t_3 )^2 + 36 t^2_4 -2 \eta^2_1 -3 \eta^2_2 
\right\},
\label{J}
\\
K
=&
\frac{8\mathbb{A}}{9} (\eta^2_1 - \eta^2_2) 
\nonumber \\
&-\frac{2\mathbb{B}}{9}
\left\{
36 t^2_2 - 12 (t_1 - t_3)^2 -36 t^2_4 +\eta^2_1 -\eta^2_2
\right\}, 
\label{K}
\\
\Gamma
=&
\frac{8 \mathbb{A}}{9} \eta^2_2 
+
\frac{\mathbb{B}}{9}
\left\{
48 (t_1 - t_3)t_2 + 72 t^2_4 - 2 \eta^2_2
\right\}, 
\label{G}
\\
\Gamma'
=&
\frac{8 \mathbb{A}}{9} \eta_1 \eta_2
-\frac{\mathbb{B}}{9} 
\left\{
24 (t_1 - t_3 -3 t_2) t_4 + 2 \eta_1 \eta_2 
\right\},
\label{G'}
\\
D
=&
-\frac{8\mathbb{A}}{9} (2t_1 + t_3) \eta_1
+\frac{16\mathbb{B}}{9} 
\left\{
(t_1 - t_3) \eta_1 - 3 t_4 \eta_2
\right\}, 
\label{D}
\\
D'
=&
-\frac{8 \mathbb{A}}{9}(2 t_1 + t_3 ) \eta_2
\nonumber \\
&-\frac{8 \mathbb{B}}{9}
\left\{
3 t_4 \eta_1 + (t_1 -t_3 + 3 t_2 ) \eta_2
\right\}, 
\label{D'} 
\end{align}
where $\mathbb{A}$ and $\mathbb{B}$ are the coefficients originating from the perturbation processes shown in Figs.~2(a) and 2(b) in the main text, respectively:
\begin{align}
\mathbb{A}
=&
-\frac{1}{3}
\,\,
\frac{J_{\rm H} + 3 \left( U + 3\lambda \right) }{6J^{2}_{\rm H} - U\left( U + 3\lambda \right) + J_{\rm H} \left( U + 4\lambda \right)} 
, \\
\mathbb{B}
=&
\frac{4}{3}
\,\,
\frac{3J_{\rm H} -U -3\lambda }{6J_{\rm H} -2U -3\lambda} \,\, 
\nonumber \\
&\times \frac{J_{\rm H}}{6J^2_{\rm H} - J_{\rm H} \left( 8U + 17\lambda \right) + \left( 2U + 3\lambda \right)\left( U + 3\lambda \right)}.
\end{align}
The contributions in Eq.~(8) in the main text are included in the terms proportional to $\mathbb{A}$ in Eqs.~\eqref{J}, \eqref{K}, and \eqref{D}.

As $J_{\rm H}$ and $\lambda$ are much smaller than $U$ in realistic situations, the coefficient $\mathbb{A}$ becomes much larger than $\mathbb{B}$; namely, $\mathbb{A}\sim1/U$ and $\mathbb{B}\sim J_{\rm H}/U^2$.
Nevertheless, the actual magnitudes as well as the signs of the exchange coupling constants are determined by the competition between the $\mathbb{A}$ and $\mathbb{B}$ terms.
For instance, in nonpolar $\alpha$-RuCl$_3$, the $\mathbb{A}$ term in the Kitaev interaction in Eq.~\eqref{K} vanishes, and the dominant FM Kitaev interaction arises from the $\mathbb{B}$ term as $K \sim-8\mathbb{B}t^{2}_2$.
However, in the polar cases discussed in the main text, the substantial increase of $\eta_1$ makes the Kitaev interaction AFM through the $\mathbb{A}$ term in Eq.~\eqref{K}.

\subsection*{S2.~~Computational details of {\it ab initio} calculations}
\begin{figure}[t]
\centering
\includegraphics[width=1\columnwidth]{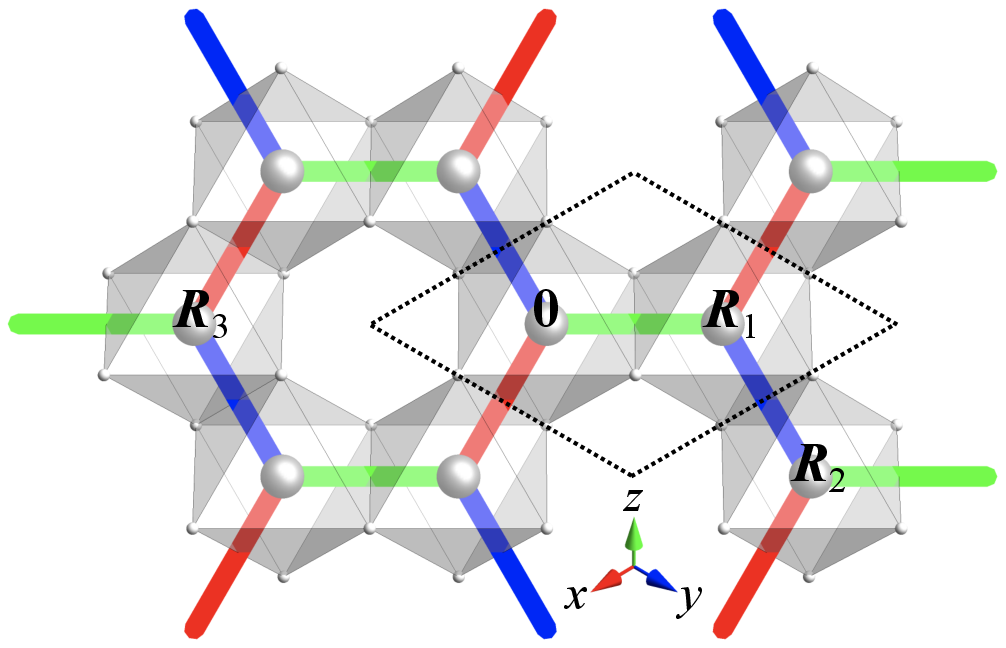}
\caption{
Atomic positions used in the estimates of transfer integrals in Table~\ref{far}.
The dotted rhombus indicates a unit cell of the monolayer honeycomb structure.
The red, blue, and green bonds for nearest neighbors denote the $x$, $y$, and $z$ bonds, respectively (see also the main text and Fig.~1).
}
\label{unit}
\end{figure}

In the {\it ab initio} calculations in the main text, we used the \textsc{openmx} code~\cite{SM.openmx}, which is based on a linear combination of pseudoatomic orbital formalism~\cite{SM.PhysRevB.67.155108,SM.PhysRevB.69.195113}.
We used the Perdew-Burke-Ernzerhof generalized gradient approximation functional in the density functional theory~\cite{SM.PhysRevLett.77.3865}, a $40\times40\times1$ $\bm{k}$-point mesh for the calculations of the self-consistent electron density and the structure relaxation, and vacuum space greater than 10~\AA~between monolayers.
In the nonrelativistic {\it ab initio} calculations, we fully optimized the primitive translational vectors and atomic positions in the unit cell (see Fig.~\ref{unit}) with the convergence criterion 10$^{-3}$ eV/\AA~for the inter-atomic forces starting from the the honeycomb structure composed of ideal octahedra; we confirmed that all the cases retain the honeycomb structures while showing trigonal distortions and tilting of ligand octahedra, which preserve the threefold rotational symmetry and the mirror symmetry discussed in the main text. 
We performed the relativistic {\it ab initio} calculations for the optimized structures.

\begin{figure}[t]
\centering
\includegraphics[width=0.72\columnwidth]{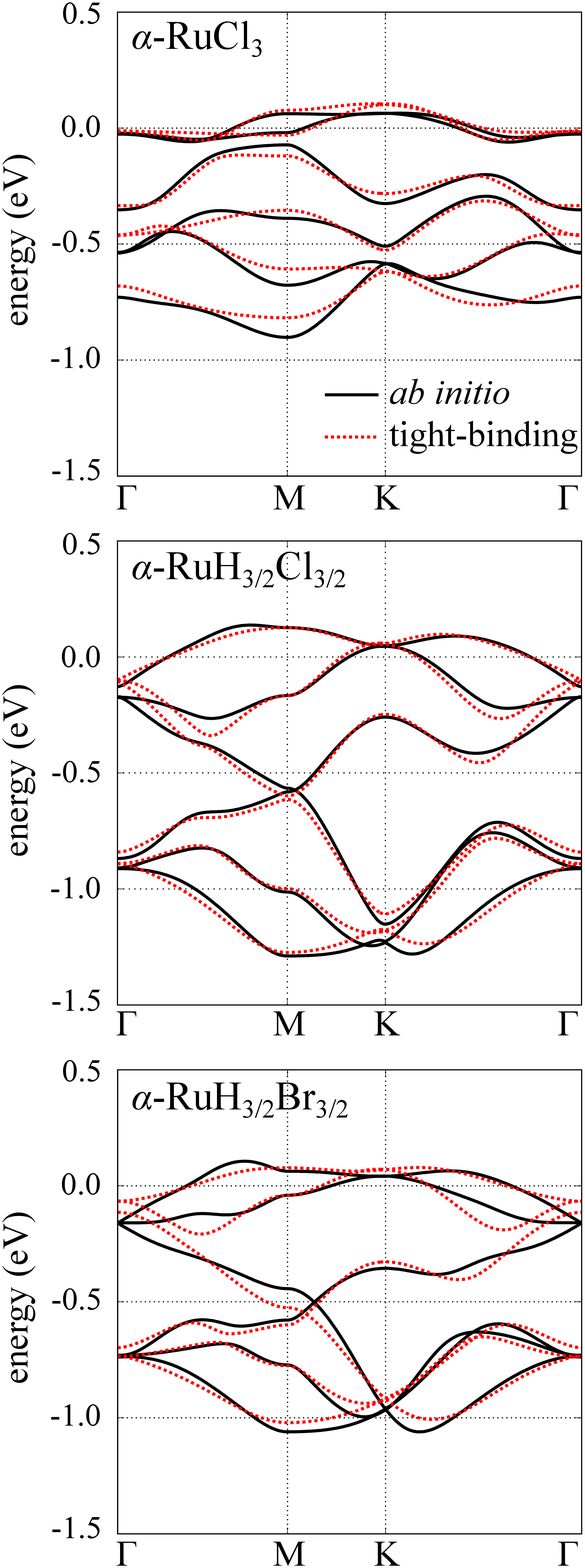}
\caption{
Electronic band structures of monolayer $\alpha$-RuCl$_3$ and $\alpha$-RuH$_{3/2}$$X_{3/2}$ ($X$= Cl and Br) in the paramagnetic states.
The Fermi level is set to zero.
The black solid lines represent the band dispersions obtained by the non-relativistic {\it ab initio} calculations, while the red dotted ones are those by the tight-binding model for the $t_{2g}$ bands with the transfer integrals and the trigonal CEF splitting estimated by the MLWF analysis. 
We take into account the transfer integrals up to third-neighbor Ru cations (see Table~\ref{far}).
}
\label{fit}
\end{figure}

To evaluate the transfer integrals and the trigonal CEF splitting, we constructed maximally-localized Wannier functions (MLWFs)~\cite{SM.PhysRevB.56.12847,SM.PhysRevB.65.035109} via a code implemented in \textsc{openmx}, for the isolated
$t_{2g}$ bands obtained by the non-relativistic calculations.
The MLWFs were stably obtained by starting from the three initial states, $d_{yz}$, $d_{zx}$, and $d_{xy}$ orbitals.
For all Ru compounds, the onsite-energy matrix for the MLWFs is well described by the form of the trigonal CEF splitting in Eq.~(5) in the main text; the deviation of the matrix element from the ideal trigonal form is typically less than 1~meV. 
Figure~\ref{fit} shows the comparison of the electronic band structures between the {\it ab initio} results and the tight-binding ones with the parameters estimated by the MLWF analysis.

\subsection*{S3.~~Transfer integrals and exchange coupling constants for further neighbors}
Table~\ref{far} summarizes the transfer integrals between the MLWFs for the Ru-Ru bonds up to third neighbors (see Fig.~\ref{unit}).
Note that the transfer integrals in Table~I in the main text correspond to those in Table~\ref{far} for the nearest-neighbor bond between the sites $\bm{0}$ and $\bm{R}_1$. 
The results for the nonpolar $\alpha$-RuCl$_3$ agree well with those in the previous work~\cite{SM.PhysRevB.93.214431}; the second-neighbor $d_{xy}$-$d_{yz}$ transfer and the third-neighbor $d_{xy}$-$d_{xy}$ transfer have relatively large amplitudes assisted by the indirect $d$-$p$-$p$-$d$ processes~\cite{SM.Winter_2017}. 
On the other hand, the polar $\alpha$-RuH$_{3/2}X_{3/2}$ ($X$= Cl and Br) possess overall larger nearest-neighbor transfers than $\alpha$-RuCl$_3$, as suggested by the optimized lattice constants discussed in the main text. 
In addition, in $\alpha$-RuH$_{3/2}X_{3/2}$, the nearest-neighbor transfers are much larger than the second- and third-neighbor ones except for the third-neighbor $d_{xy}$-$d_{xy}$ transfer.
The large third-neighbor $d_{xy}$-$d_{xy}$ transfer also would be ascribed to the compression of the lattice constants, which enhances both direct $d$-$d$ and indirect $d$-$p$-$p$-$d$ transfers. 
We note that the tight-binding model constructed from these nearest-, second-, and third-neighbor transfers well reproduce the {\it ab initio} band structures (see Fig.~\ref{fit}). 

\begin{table}
\begin{ruledtabular}
\begin{tabular}{c|ccc}
$\alpha$-RuCl$_3$				&$\bm{R}_{1}$		&$\bm{R}_{2}$		&$\bm{R}_{3}$\\
\hline
($yz$, $yz$)					&$45$			&$-3$			&$-9$\\
($yz$, $zx$)					&$159$			&$-2$			&$-8$\\
($yz$, $xy$)					&$-22$			&$-25$			&$13$\\
($zx$, $yz$)					&$159$			&$9$				&$-8$\\
($zx$, $zx$)					&$45$			&$-2$			&$-9$\\
($zx$, $xy$)					&$-22$			&$-2$			&$13$\\
($xy$, $yz$)					&$-22$			&$-59$			&$13$\\
($xy$, $zx$)					&$-22$			&$9$				&$13$\\
($xy$, $xy$)					&$-117$			&$-3$			&$-42$\\
\hline
\hline
$\alpha$-RuH$_{3/2}$Cl$_{3/2}$	&$\bm{R}_{1}$		&$\bm{R}_{2}$		&$\bm{R}_{3}$\\
\hline
($yz$, $yz$)					&$143$			&$26$			&$1$\\
($yz$, $zx$)					&$-177$			&$-8$			&$-18$\\
($yz$, $xy$)					&$-66$			&$-2$			&$30$\\
($zx$, $yz$)					&$126$			&$9$				&$27$\\
($zx$, $zx$)					&$143$			&$10$			&$1$\\
($zx$, $xy$)					&$-96$			&$18$			&$8$\\
($xy$, $yz$)					&$-96$			&$-34$			&$8$\\
($xy$, $zx$)					&$-66$			&$-35$			&$30$\\
($xy$, $xy$)					&$-227$			&$2$				&$-144$\\
\hline
\hline
$\alpha$-RuH$_{3/2}$Br$_{3/2}$	&$\bm{R}_{1}$		&$\bm{R}_{2}$		&$\bm{R}_{3}$\\
\hline
($yz$, $yz$)					&$97$			&$22$			&$-1$\\
($yz$, $zx$)					&$-142$			&$-7$			&$-16$\\
($yz$, $xy$)					&$-55$			&$-18$			&$25$\\
($zx$, $yz$)					&$141$			&$12$			&$15$\\
($zx$, $zx$)					&$97$			&$7$				&$-1$\\
($zx$, $xy$)					&$-79$			&$-7$			&$13$\\
($xy$, $yz$)					&$-79$			&$-44$			&$13$\\
($xy$ ,$zx$)					&$-55$			&$40$			&$25$\\
($xy$, $xy$)					&$-128$			&$0$				&$-161$\\
\end{tabular}
\end{ruledtabular}
\caption{
Transfer integrals between the MLWFs for the $t_{2g}$ bands of monolayer $\alpha$-RuCl$_3$ and $\alpha$-RuH$_{3/2} X_{3/2}$ ($X$= Cl and Br).
Each value means $\bra{d_m,\bm{0}} H \ket{d_n,\bm{r}}$, where $H$ is the Hamiltonian of the system and $\ket{d_m,\bm{r}}$ is the $d_{m}$-like MLWF at site $\bm{r}$ ($m$ = $yz$, $zx$, and $xy$).
See Fig.~\ref{unit} for the spatial positions of $\bm{r}$.
The unit is in meV.
}
\label{far}
\end{table}

From the estimates of the transfer integrals, we also evaluate the exchange coupling constants between the pseudospins up to third-neighbor bonds (see Sec.~S4 for the details of the calculations). 
For all compounds in the present study, we find that  the second-neighbor exchange couplings are smaller than 1~meV.
We present the exchange coupling constants for the third neighbor in Table~\ref{3coupling}.
From the symmetry, the exchange interaction for the third neighbor, $\bm{S}^{\rm T}_{\bm{0}} \hat{J}^{(3)} \bm{S}_{\bm{R}_3}$, can be represented in the same manner as that for the nearest neighbor in Eq.~(6) in the main text as
\begin{equation}
\hat{J}_{z}^{(3)}
=
\left(
\begin{array}{ccc}
J_3				&D_3 + \Gamma_3		&-D'_3 + \Gamma'_3\\
-D_3 + \Gamma_3	&J_3 				&D'_3 + \Gamma'_3\\
D'_3 + \Gamma'_3	&-D'_3 + \Gamma'_3		&J_3 + K_3
\end{array}
\right).
\end{equation}
For $\alpha$-RuH$_{3/2}X_{3/2}$, while several exchange coupling constants have amplitudes of a few meV due to the relatively large $d_{xy}$-$d_{xy}$ transfer integral, they are much smaller than the Heisenberg and Kitaev interactions for the nearest neighbors (see Table~II in the main text). 

\begin{table}
\centering
\begin{ruledtabular}
\begin{tabular}{c|cccccc}
							&$J_3$	&$K_3$	&$\Gamma_3$	&$\Gamma'_3$	&$D_3$	&$D'_3$\\
\hline
$\alpha$-RuCl$_{3}$				&$0.9$	&$0.2$	&$0.0$		&$-0.2$		&$0.0$	&$0.0$\\
$\alpha$-RuH$_{3/2}$Cl$_{3/2}$	&$-3.2$	&$3.9$	&$0.3$		&$0.1$		&$4.2$	&$1.7$\\
$\alpha$-RuH$_{3/2}$Br$_{3/2}$	&$3.9$	&$4.3$	&$0.5$		&$-0.1$		&$3.0$	&$0.7$	
\end{tabular}
\end{ruledtabular}
\caption{
Exchange coupling constants for the third neighbor in monolayers of nonpolar $\alpha$-RuCl$_3$ and polar $\alpha$-RuH$_{3/2}X_{3/2}$ ($X$= Cl and Br) estimated by using the parameters in Table~\ref{far} and the empirical values of the Coulomb interactions and the SOC for Ru$^{+3}$: $U=3.0$~eV, $J_{\rm H}=0.6$~eV, and $\lambda=0.15$~eV~\cite{SM.PhysRevB.93.214431}.
The unit is in meV.
}
\label{3coupling}
\end{table}

\subsection*{S4.~~Second-order perturbation}

To evaluate the exchange coupling constants based on the second-order perturbation theory,  we need to choose the proper pseudospin basis describing the lowest-energy Kramers pair of holes in the $t_{2g}$ manifold in the presence of both
SOC and trigonal CEF splitting [see Eqs.~(4) and (5) in the main text].
By considering the mirror symmetry of the system with respect to the plane spanned by the [001] and [110] axes through the center of nearest-neighbor bonds, we take the quantization axis of the pseudospins on this plane.
In addition, by taking into account the threefold rotational symmetry around the [111] axis of the system, we employ the pseudospin states that are transformed by the threefold rotations around the [111] axis in the same manner as the SU(2) spin. Then, we obtain the basis as
\begin{align}
\ket{\tilde{\uparrow}}
&=
\frac{s_{\theta}}{\sqrt{6}} 
\ket{d_{yz} \uparrow}
+
(\frac{c_{\theta}}{\sqrt{3}} + {\rm i}\frac{s_{\theta}}{\sqrt{6}})
\ket{d_{yz} \downarrow}
+
\frac{s_{\theta}}{\sqrt{6}}
\ket{d_{zx} \uparrow}
\nonumber \\
&+
(\frac{s_{\theta}}{\sqrt{6}} + {\rm i}\frac{c_{\theta}}{\sqrt{3}})
\ket{d_{zx} \downarrow}
+
\frac{c_{\theta}}{\sqrt{3}}
\ket{d_{xy} \uparrow}
+
\frac{s_{\theta}}{\sqrt{3}} e^{{\rm i} \frac{\pi}{4}}
\ket{d_{xy} \downarrow},  
\label{up}
\\
\ket{\tilde{\downarrow}}
&=
(\frac{c_{\theta}}{\sqrt{3}} - {\rm i}\frac{s_{\theta}}{\sqrt{6}})
\ket{d_{yz} \uparrow}
-
\frac{s_{\theta}}{\sqrt{6}}
\ket{d_{yz} \downarrow}
+
(\frac{s_{\theta}}{\sqrt{6}} -{\rm i} \frac{c_{\theta}}{\sqrt{3}})
\ket{d_{zx} \uparrow}
\nonumber \\
&-
\frac{s_{\theta}}{\sqrt{6}}
\ket{d_{zx} \downarrow}
+
\frac{s_{\theta}}{\sqrt{3}} e^{-{\rm i} \frac{\pi}{4}}
\ket{d_{xy} \uparrow}
-
\frac{c_{\theta}}{\sqrt{3}}
\ket{d_{xy} \downarrow},    
\label{down}
\end{align}
where $c_{\theta}=\cos{\theta}$, $s_{\theta}=\sin{\theta}$, $\theta = \frac12 \arctan[4\sqrt{2} \Delta_{\rm tri}/(3\lambda - 2\Delta_{\rm tri})]$, and $\ket{d_m \, \sigma}$ denotes the single-hole state with orbital $d_m$ ($m=yz$, $zx$, or $xy$) and spin $\sigma=\uparrow$ or $\downarrow$  (the spin quantization axis is taken along the [001] axis)~\cite{SM.khaliullin2005orbital}. 
By using this basis, the effective total angular momentum $j_\gamma$ ($\gamma=x$, $y$, and $z$) of the pseudospin is given as 
\begin{equation}
(\bra{\tilde{\uparrow}} \, \bra{\tilde{\downarrow}})^{\rm T}
j_{\gamma}
(\ket{\tilde{\uparrow}} \, \ket{\tilde{\downarrow}})
=
\frac{\cos 2\theta}{2} \hat{\sigma}_\gamma
+\frac{\sin^2 \theta}{2} (\hat{\sigma}_{\bar{\gamma}} + \hat{\sigma}_{\bar{\bar{\gamma}}}),
\end{equation}
where $\left(\gamma, \bar{\gamma}, \bar{\bar{\gamma}} \right)$ denotes a cyclic permutation of $\left(x, y, z \right)$.
We note that these expressions give the $j_{\rm eff}=1/2$ states in the limit of $\lambda \gg \Delta_{\rm tri}$~\cite{SM.PhysRevLett.102.017205}. 

In the ground state, there is a single hole in the $t_{2g}$ manifold at each site.
Then, we can evaluate the exchange interactions between the pseudospins at sites $i$ and $j$ from the second-order energy corrections to the ground-state energy $E_0$:
\begin{align}
&E^{(2)}_{\tilde{\sigma}'_i, \tilde{\sigma}'_j;\,\tilde{\sigma}_i, \tilde{\sigma}_j} 
=
\nonumber \\
&
\bra{\tilde{\sigma}'_i}
\bra{\tilde{\sigma}'_j}
\,
H_{\rm hop}
\,\,
\frac{1}{E_0 - (H_{\rm SOC} + H_{\rm tri} + H_{\rm int})}
\,\,
H_{\rm hop}
\,
\ket{\tilde{\sigma}_i}
\ket{\tilde{\sigma}_j}
,
\end{align}
where $\tilde{\sigma}_{i}$ ($\tilde{\sigma}'_{i}$) denotes the pseudospin $\tilde{\uparrow}$ or $\tilde{\downarrow}$ at site $i$ in the initial (final) state in the perturbation process~\cite{SM.PhysRevLett.113.107201}. 
We confirm that in the absence of the trigonal CEF splitting, these estimates give consistent values with the analytical expressions in Eqs.~\eqref{J}-\eqref{D'}. 


\end{document}